\documentclass{elsart4}

\usepackage{amssymb}

\usepackage{graphics}
\usepackage{graphicx}
\usepackage{epsfig}
\usepackage{amssymb}
\journal{Physica B}

\begin{document}

\begin{frontmatter}


 \title{Wigner crystal in snaked nanochannels: outlook}
 \author{O.V.Zhirov }
 \address{Budker Institute of Nuclear Physics, 630090 Novosibirsk, Russia}
 \author{D.L.Shepelyansky}
 \address{Laboratoire de Physique Th\'eorique du CNRS (IRSAMC), 
Universit\'e de Toulouse, UPS, F-31062 Toulouse, France}
\ead{dima@irsamc.ups-tlse.fr}





\begin{abstract}
We study properties of Wigner crystal in snaked nanochannels and
show that they are characterized by a conducting sliding phase
at low charge densities and an insulating pinned phase 
emerging above a certain critical
charge density. We trace parallels between this model problem and 
the Little suggestion for electron transport in organic molecules.
We also show that in presence of periodic potential inside the 
snaked channel the sliding phase exists only inside a certain window of 
electron densities that has similarities with a pressure dependence of 
conductivity in organic  conductors. Our studies show emergence of 
dynamical glassy phase in a purely periodic potential in absence of
any disorder that can explain enormously slow variations of 
resistivity in organic conductors. Finally we discuss 
the KAM concept of superfluidity induced by repulsive Coulomb interaction
between electrons. We argue that the transition
from the sliding KAM phase to the pinned Aubry phase corresponds to 
the superfluid-insulator transition. 
\end{abstract}

\begin{keyword}
     Wigner crystal, organic conductors, Frenkel-Kontorova model, Aubry transition, invariant KAM curves
\PACS 05.45.Ac, 71.45.Lr, 82.47.Uv  
\end{keyword}
\end{frontmatter}

\section{Introduction}
\label{sect1}
\begin{figure}
\includegraphics[width=7.6cm]{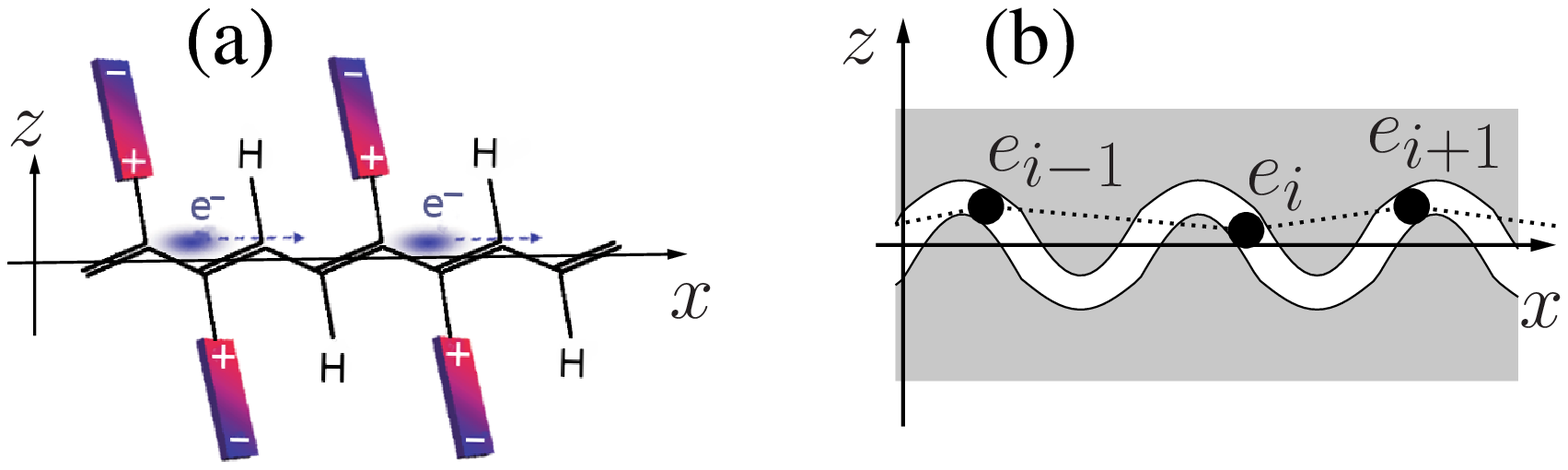}
\caption{  }
\label{fig1}
\end{figure}

The Wigner crystal \cite{wigner} appears when the energy 
of Coulomb repulsion between
charges of same sign becomes dominant comparing 
to kinetic energy of charge motion.
On a one-dimensional (1D) straight line this crystal can move ballistically
as a whole at an arbitrary small velocity.
Here,  we discuss the properties of Wigner crystal sliding in 1d 
in presence of a periodic potential and in 1d snaked nanochannel 
following recent works \cite{fki,snake}.
An example of snaked nanochannel of a sinusoidal form 
is shown in Fig.~\ref{fig1}.
The snaked form of a channel is very similar to
the Little suggestion \cite{little} on possibilities of superconductivity in
organic molecules. As described in \cite{little,jerome},
it is assumed that organic molecules form some effective 
wiggled or snaked channel with an effective density of electrons $\nu$
which slides along the channel opening a new view on possibilities 
of superconductivity in such materials.

 The question about sliding in such a channel is rather nontrivial
being linked with fundamental results of dynamical systems
\cite{chirikov,lichtenberg} which we briefly discuss. 
In fact in a local approximation of small charge oscillations
the forces depend linearly on displacements
corresponding to a string of particles, linked 
 locally by linear springs
and placed in a periodic
potential. The density of particles or charges 
corresponds effectively to a fixed rotation number in 
a dynamical symplectic map which describes the recurrent
positions of particles in a static configuration with a minimum of energy.
For linear springs this map is exactly reduced to the Chirikov standard map
\cite{chirikov}. 
This model in known also as the Frenkel-Kontorova model
which detailed description is given in \cite{obraun}.
At small channel deformation $a$
or small amplitude of periodic potential $K$
the particles can freely slide in the periodic potential
that corresponds to the regime of invariant 
Kolmogorov-Arnold-Moser (KAM) curves which 
rotation number determines the particle density.
In this regime the spectrum of small excitations
is characterized by a phonon spectrum
with the dispersion relation $\omega=c_s k$
where $k$ is dimensionless wave vector and $c_s$
is dimensionless sound velocity.
Above a certain critical strength of deformation
the KAM curve at a given $\nu$ is destroyed
being replaced by an invariant Cantor set known as 
cantori \cite{aubry}. In this regime the excitations
above the ground state have a gap
$\omega^2=(c_sk)^2+\Delta^2$ and the chain becomes pinned by the potential.
The gap $\Delta$ is proportional to the Lyapunov exponent
of dynamical orbits on such cantori set \cite{lichtenberg,aubry}.
The transition between sliding and pinned phases is known as 
the Aubry transition \cite{obraun}.
In the pinned phase the Aubry theorem guarantees that at fixed $\nu$ there is 
a unique ground state which static equilibrium configuration
corresponds to the cantori set with positions of particles
forming a devil's staircase. However, from the physical view point this 
ground state is rather hard to reach since
in its vicinity there are exponentially many
equilibrium configurations which energy is exponentially
close to the energy of the ground state.
Numerical studies show that already for 100 particles
 the energy difference is as small as $10^{-25}$
in dimensional units \cite{fk1}. This phase was called the 
dynamical spin glass (or dynamical instanton glass)
since such properties appear in spin glasses \cite{parisi}
which have random disordered on-site energies and interactions.
In contrast to that these properties of the Aubry phase
appear in absence of any disorder being of purely dynamical origin
of cantori in a purely periodic potential. 

The studies of properties of the quantum Frenkel-Kontorova
model has been started in \cite{fkq1} and further 
significantly advanced  in 
\cite{fkq2}. It was shown \cite{fkq2}
that quantum fluctuations lead to melting of 
the pinned phase at sufficiently 
large values of dimensionless Planck constant $\hbar$.
This transition is a zero temperature $T=0$ quantum phase transition.
At small  $\hbar \ll 1$ an $T=0$ the phonon mode
is frozen but quantum tunneling gives transitions
between quasi-degenerate
equilibrium classical configurations which can be viewed 
as instantons. At small $\hbar \ll 1$
the density of instantons is small
and their interactions are weak.
When $\hbar$ increases
the instanton density grows and
above a certain critical $\hbar_c \sim 1$
the quantum melting of pinned phase
takes place at zero temperature 
leading to zero gap, appearance of quantum 
phonon mode and quantum chain sliding \cite{fkq2}.
The results obtained for the Wigner crystal in a periodic potential
\cite{fki} in classical and quantum regimes
confirm this qualitative picture.
At fixed amplitude of periodic potential
$K$ the classical Wigner crystal is pinned at
small charge densities $\nu < \nu_{c1}$ \cite{fki}.
Indeed, at $\nu \rightarrow 0$ we have a problem of one electron
with zero kinetic energy and obviously, 
an electron is pinned by a periodic potential.

\section{Sliding Wigner snake}
\label{sect2}

The situation is different in the case of
snaked nanochannel: noninteracting electrons, 
corresponding to the limit $\nu \ll 1$,
move freely inside the wiggled channel
and pinning of the Wigner crystal 
appears only above a certain critical 
charge density $\nu > \nu_2$.
An example of sliding and pinned regimes
is shown in Fig.2. The data clearly show that
the sliding phase 
at $a=0.6$ has a smooth hull function
and sound dispersion law for small oscillations
of the crystal. In contrast,
in the pinned regime at $a=1.2$
the hull function has a form of fractal devil's staircase
and gapped spectrum of small oscillations.  
\begin{figure}
\includegraphics[width=7.6cm]{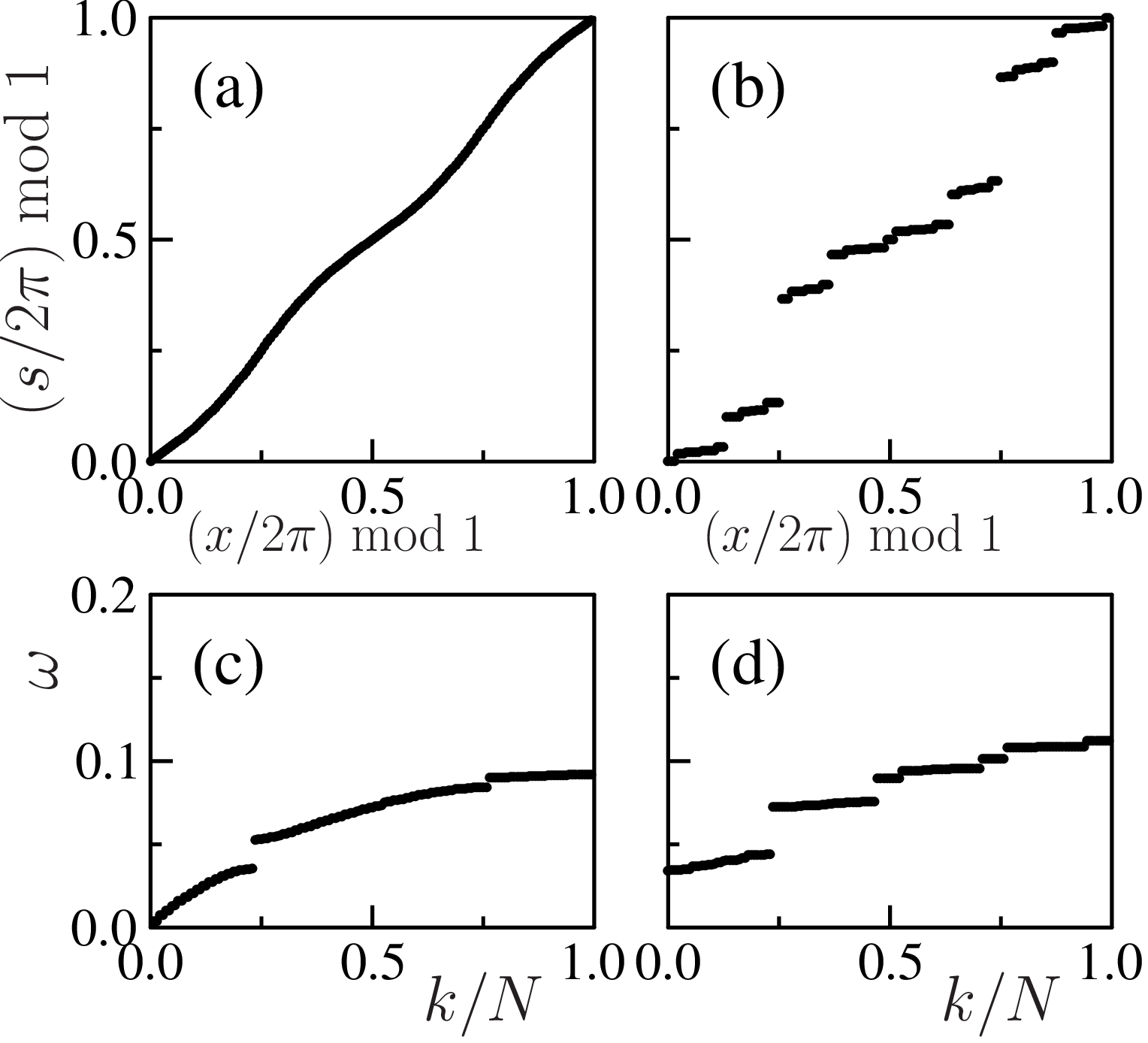}
\caption{
 }
\label{fig2}
\end{figure}

\begin{figure}
\includegraphics[width=7.6cm]{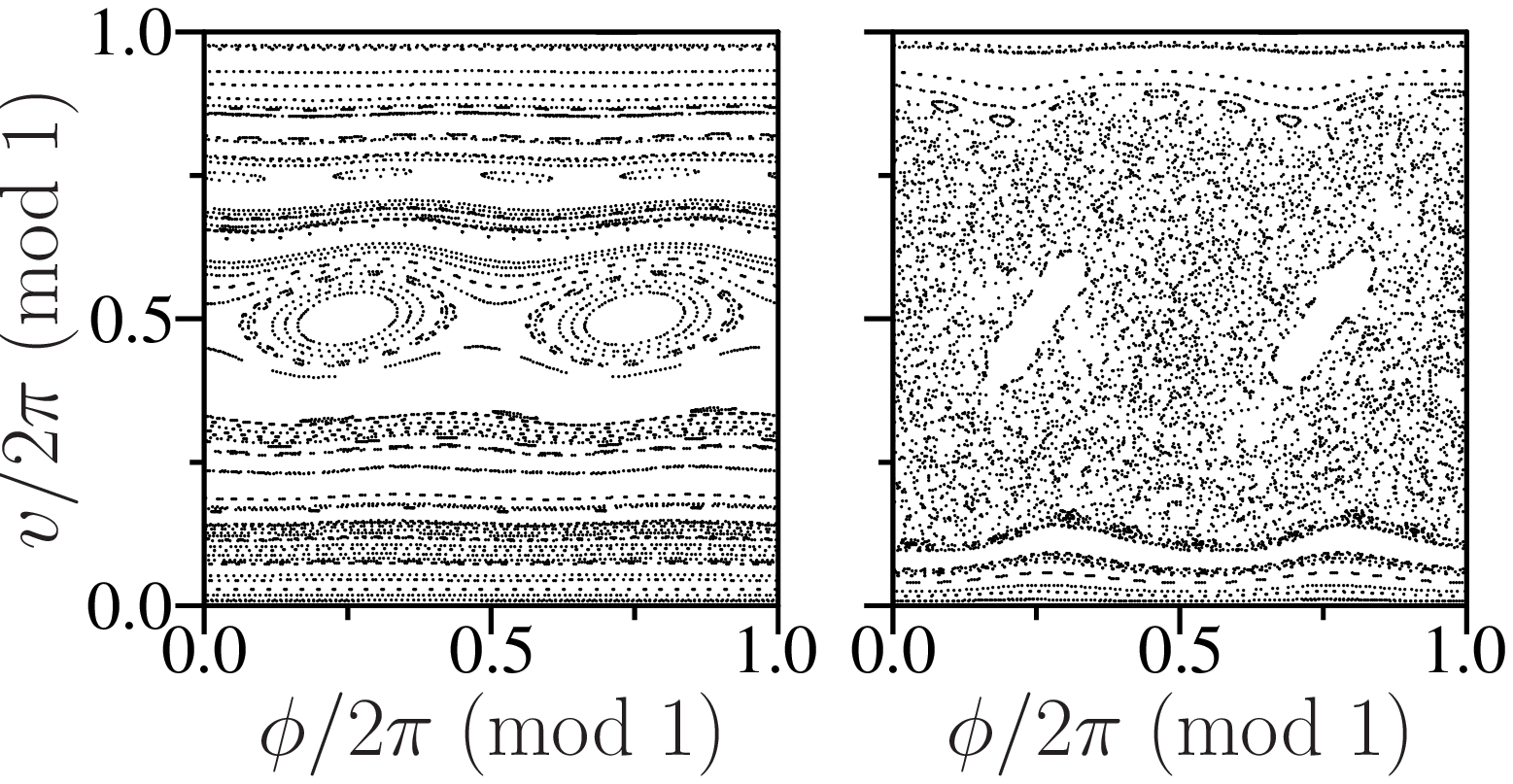}
\caption{
} 
\label{fig3}
\end{figure}

More detailed results on dependence of
gap $\Delta$ on charge density $\nu$
and deformation $a$ are described in \cite{snake}.
In \cite{snake} it is also shown that for moderate
deformations $a <1$ the charge positions
in a static configuration
are described by a symplectic dynamical map
\begin{eqnarray}
\label{eq1}
 {\bar v} &  = & v + 2 a^2 (1-\cos {\bar v}) \sin 2 \phi \; ,\nonumber\\
{\bar \phi} & = & \phi + {\bar v} + a^2 \sin {\bar v} \cos 2 \phi \; ,
\end{eqnarray}
where $v=x_{i}-x_{i-1}$, $\phi=x_i$ are conjugated action-phase variables,
bar marks their values after iteration.
The map is implicit but symplectic (see e.g. \cite{lichtenberg}).
Examples of Poincar\'e sections of this map
at two values of deformation $a$ are shown in
Fig.~\ref{fig3}. Phase space region with scattered points
corresponds to
chaotic dynamics with pinned phase, while the smooth 
invariant KAM curves correspond to the sliding phase.

In many aspects the properties of 
the Wigner crystal in snaked nanochannels 
are similar to those of the Frenkel-Kontorov model
\cite{fk1}
and the Wigner crystal in a periodic potential \cite{fki}:
in the pinned phase there are exponentially
many static configurations being
exponentially close in energy and corresponding to
the dynamical glass phase. However, there are also some specific features: for
rational values of densities $\nu=\nu_m=1/m$,
where $m$ is an integer, the Wigner
snake can slide freely since a displacement does not modify
the Coulomb energy of electron interactions.

In analogy with the results presented in \cite{fki},
we expect that the quantum Wigner crystal 
shows a zero temperature quantum phase transition
going from a pinned phase at $\hbar <\hbar_c \sim 1$
to a sliding phase at   $\hbar > \hbar_c $.
However, a direct demonstration of this fact
requires further numerical simulations using 
quantum Monte Carlo methods described in \cite{fki,fkq1,fkq2}.

\section{Discussion}
\label{sect3}

\begin{figure}
\includegraphics[width=7.4cm]{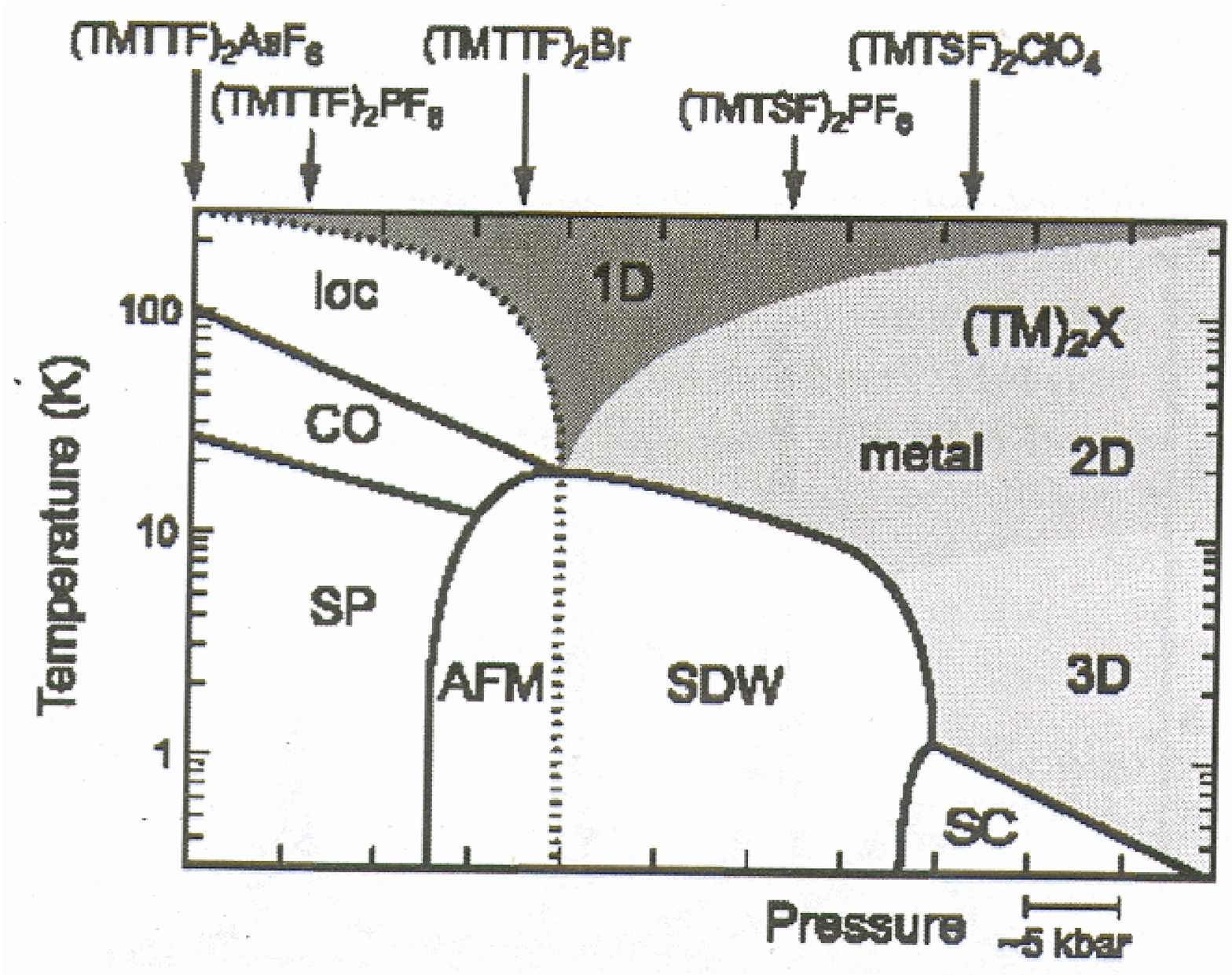}
\caption{
} 
\label{fig4}
\end{figure}

In the above Section we considered the Wigner crystal in 
a snaked nanochannel without any internal potential.
It is natural to assume that a more realistic case of molecular 
organic conductors, as shown in the Little suggestion in Fig.~\ref{fig1},
has not only channel deformation but also 
a periodic potential inside the channel.
Thus the case of organic conductors corresponds to
a case of snaked channel with a periodic potential inside it.
The combination of results of \cite{fki,snake}
shows that for a given deformation and amplitude of the periodic potential
we have the sliding phase in a certain
range of charge densities $\nu$:
\begin{equation}
\nu_{c1} < \nu <\nu_{c2}   \;\;\; .
\label{eq2}
\end{equation}
We suppose that the sliding KAM phase 
may correspond to effective superconducting behavior of 
electron transport in organic conductors.
Indeed, the pressure diagram of organic conductors
shown in Fig.~\ref{fig4} from \cite{brazovski}
shows that superconductivity exists only
is a finite range of pressure.
We assume that pressure gives variation of 
effective charge density inside the 
molecular channels in the Little suggestion 
in Fig.~\ref{fig1}.
This leads us to the KAM concept of 
superconductivity of electrons without attractive forces:
the Wigner crystal of electrons slides freely
inside a snaked molecular crystal channel
if the charge density is located inside of
KAM phase defined by (\ref{eq2}).
Of course, further studies are required for
development of this concept. 
In fact. the sliding KAM phase
can be viewed as a superfluid phase of electrons.
Indeed, we see that in the KAM phase
there is a  spectrum of excitations
with a finite sound velocity $c_s$.
Thus according to the Landau criterion \cite{landau}
the sliding of electrons 
with velocities $v<c_s$ is superfluid.
Hence, the transition from the sliding KAM
phase to the pinned Aubry phase corresponds to
the transition from superfluid to insulator.
In this superfluid liquid
the charge carries have change ``e'' and not
``2e'' as it is the case for BCS pairs.
May be effect of interactions between electrons in
parallel snaked channels should be taken into
account to have 2e-pairs. We note that it is known that repulsive
interactions can create superfluid phase in
disordered 1d systems, e.g. in the repulsive Hubbard model
with disorder \cite{scalettar}.

The existence of dynamical spin glass phase
with pinned Wigner crystal shows that
there should be very slow relaxation processes
corresponding to very slow transitions between quasi-degenerate
static equilibrium
configurations. In fact the experiments with 
organic conductors show very slow variations of conductivity
which take place of a scale of days.
Such experimental results have been reported at ECRYS-2011
by K.Miyagawa \cite{miyagawa} and P.Monceau \cite{monceau}. 
Usually it is argued that the glassy phase appears due to impurities.
We think that the origin of this phenomenon is not
related to disorder and impurities, which presence should be 
rather small  in organic crystals used in experiments
\cite{miyagawa,monceau}. In contrast this glassy phase appears as a result
of dynamical spin glass phase described in \cite{fki,fkq1,fkq2,snake}
which exists in purely periodic structures without any
impurities and disorder.

Finally, we note that in \cite{fki} it was proposed to study 
the dynamical spin glass with cold ions in optical lattices
which can model the problem of Wigner crystal in
a periodic potential. Such experiments with cold ions
are now under active discussions \cite{wunderlich,tosatti}
and their experimental realization is on the way \cite{haffner}.

This work is supported in part by 
ANR PNANO project NANOTERRA.

\newpage
Figure captions

Fig.1 (a)A schematic image of the Little suggestion
for electron transport in organic molecules (after \cite{little,jerome}).
(b) A schematic image of electron Wigner crystal 
with charges $e_i$ (points) 
sliding in a snaked sinusoidal nanochannel,
dashed lines show force directions between nearby electrons.

Fig.2 Hull function $s=h(x)$ (a,b)
and phonon spectrum $\omega(k/N)$ (c,d) 
for incommensurate electron
density $\nu=N/L=144/233$ is shown
at $a=0.6$ (a,c) and $a=1.2$ (b,d).
Here $x$ gives the positions $s_i$ of electrons at
$a=0$. Here $a$ gives the dimensional amplitude of 
sinusoidal channel described by 
equation $y=a\sin x$.

Fig.3 Poincar\'e section for the dynamical map
(\ref{eq1}) 
at $a=0.2$ (left panel), $0.6$ (right panel).

Fig.4 Schematic phase diagram of the $(TMTCF)_2X$ family
taken from \cite{brazovski} (see details there).


\begin{thebibliography}{00}

\bibitem{wigner} E.~Wigner, Phys. Rev. {\bf 46} (1934) 1002.
\bibitem{fki} I.~Garcia-Mata, O.V.~Zhirov and D.L.~Shepelyansky,
        Eur. Phys. J. D {\bf 41} (2007) 325.
\bibitem{snake} O.V.Zhirov and D.L.Shepelyansky, 
        Eur. Phys. J. B {\bf 82} (2011) 61.
\bibitem{little} W.A.~Little, Phys. Rev. A {\bf 134} (1964) 1416;
                 Sci. Am. {\bf 212} (1965) 21.
\bibitem{jerome} D.~J\'erome,  ``Historical approach to organic\\ 
        superconductivity'',  in 
        {\it The Physics} {\it of Organic Superconductors}\\ 
        {\it and Conductors}, A.Lebed (Ed.),
        p.3, Springer-Verlag, Berlin (2008).
\bibitem{chirikov} B.V.~Chirikov, Phys. Rep. {\bf 52} (1979) 263.
\bibitem{lichtenberg} A.J.~Lichtenberg and M.A.~Lieberman, 
        {\it Regular and chaotic dynamics}, Springer, Berlin (1992).
\bibitem{obraun} O.M.~Braun and Yu.S.~Kivshar, 
        {\it The Frenkel-Kontorova Model: Concepts, Methods, Applications}, 
        Springer-Verlag, Berlin (2004).
\bibitem{aubry} S.~Aubry, Physica D {\bf 7} (1983) 240.
\bibitem{fk1} O.V.Zhirov, G.Casati and D.L.Shepelyansky, Phys. Rev. E
        {\bf 65} (2002) 026220.
\bibitem{parisi} M.~M\'ezard, G.~Parisi and M.A.~Virasoro,
        {\it Spin Glass Theory and Beyond},
        World Sci., Singapore (1987).
\bibitem{fkq1} F.Borgonovi, I.Guarneri and D.L.Shepelyansky,
         Phys. Rev. Lett. {\bf 63} (1989) 2010.
\bibitem{fkq2}  O.V.Zhirov, G.Casati and D.L.Shepelyansky, Phys. Rev. E
        {\bf 67} (2003) 056209.
\bibitem{brazovski} S.A.Brazovskii, ``Ferroelectricity and 
        charge ordering in quasi-1D organic conductors'', in 
        {\it The Physics} {\it of Organic Superconductors and Conductors} 
        A.Lebed (Ed.),
        p.313, Springer-Verlag, Berlin (2008).
\bibitem{landau} L.Landau, Phys. Rev. {\bf 60(4)}, 356 (1941).
\bibitem{scalettar} R.T.Scalettar, G.G.Batrouni, and G.T.Zimanyi,
        Phys. Rev. Lett.   {\bf 66}, 3144 (1991).
\bibitem{miyagawa} K.Miyagawa, Y.Muto, M.Kodama,
        Y.Shimada and K.Kanoda,
        ``$^{13}C$ NMR and resistivity measurements of charge glass state 
         in a charge-frustrated organic conductor,
         $q-(BEDT-TTF)_2RNZN(SCN)_4$'',
          report at ECRYS-2011, Cargese (see this Physica B issue)
\bibitem{monceau} P.Monceau, ``Ferroelectricity in organic low-dimensional
        systems'', report at ECRYS-2011, Cargese (see this Physica B issue)
\bibitem{wunderlich} M.Johanning, A.F.Varon and C.Wunderlich,
        J. Phys. B: At. Mol. Opt. Phys. {\bf 42} (2009) 154009.
\bibitem{tosatti} A.Benassi, A.Vanossi and E.Tosatti,
        Nature Comm. {\bf 2} (2011) 236.
\bibitem{haffner} T.Pruttivarasin, M.Ramm, I.Talukdar, A.Kreuter 
        and H.H\"affner, New J. Phys. {\bf 13} (2011) 075012.

\end{thebibliography}
\end{document}